\documentclass[onecolumn]{elsart}
\usepackage{epsfig}
\usepackage{amssymb}
\usepackage{amsmath}
\usepackage{float}
\usepackage{subfigure}

\newcommand{\cal}{\mathcal}

\newcommand{\ep}{\varepsilon}

\newcommand{\eqs}[1]{\begin{equation} \begin{split} #1\end{split} \end{equation} }

\newcommand{\ce}[1]{Eq.~(\ref{#1})}

\newcommand{\cf}[1]{{Fig.~\ref{#1}}}

\begin{document}

\begin{frontmatter}

\title{Hadroproduction of $J/\psi$ and $\Upsilon$ \\
in association with a heavy-quark pair}

\author[UCL]{P.~Artoisenet},
\author[X]{J.P.~Lansberg},
\author[UCL]{F.~Maltoni}

\address[UCL]{Center for Particle Physics and Phenomenology (CP3), \\
Universit\'e catholique de Louvain, B-1348 Louvain-la-Neuve, Belgium}
\address[X]{CPHT, Ecole Polytechnique, CNRS, 91128 Palaiseau Cedex, France}

\begin{abstract}
We calculate the cross sections for the direct hadroproduction of
$J/\psi$ and $\Upsilon$ associated with a heavy-quark pair of the same
flavour at leading order in $\alpha_S$ and $v$ in NRQCD.  These
processes provide an interesting signature that could be studied at
the Tevatron and the LHC and also constitute a gauge-invariant subset
of the NLO corrections to the inclusive hadroproduction of $J/\psi$
and $\Upsilon$.  We find that the fragmentation approximation commonly
used to evaluate the contribution of these processes to the inclusive
quarkonium production sizeably underestimates the exact calculation in
the kinematical region accessible at the Tevatron.  Both 
$J/\psi$ and $\Upsilon$ are predicted to be unpolarised, 
independently of their transverse momentum.

\end{abstract}

\end{frontmatter}

\section{Introduction}

Since its introduction in 1995~\cite{Bodwin:1994jh,yr}, NRQCD has
become the standard framework to study heavy-quarkonium physics.
NRQCD is a non-relativistic effective theory equivalent to QCD, where
inclusive cross sections and decay rates can be factorised as the product
of short-distance and long-distance parts.  The short-distance
coefficients are process-dependent but can be computed in perturbative
QCD, the strong coupling constant $\alpha_S$ being the parameter of
the expansion. The long-distance matrix elements are
process-independent, {\it i.e.}, universal, but non-perturbative. They can be
classified in terms of their scaling in $v$, the relative velocity
of the heavy quarks in the bound state, and as a result, physical
quantities can be expressed as a double series in $\alpha_S$ and $v$.

An innovation of this formalism is the colour-octet mechanism: the
heavy-quark pair is allowed to be created in a colour-octet state over
short distances, the color being neutralised over long distances.  It
is thanks to this very mechanism that it is possible to account for
the CDF data~\cite{Abe:1997jz,Abe:1997yz} on the inclusive $J/\psi$
and $\psi'$ cross sections at the
Tevatron~\cite{Braaten:1994vv,Cho:1995vh,Cho:1995ce}. In this case the
colour-octet matrix elements fitted from the data roughly scale as the
power counting rules of NRQCD predict. On the other hand, the recent
data collected at $\sqrt{s}=1.96$ TeV by the CDF
collaboration~\cite{Affolder:2000nn} have revealed that the $J/\psi$
is unpolarised, in flagrant disagreement with the expectations of
NRQCD. It is fair to say that the mechanisms responsible for the
quarkonium production at the Tevatron are not completely understood
yet (for a recent review see~\cite{Lansberg:2006dh}).

Another challenge to theorists has been provided by the recent
measurements at $B$ factories, where rates for inclusive and exclusive
$J/\psi$ production in association with a $c\bar{c}$ quark
pair~\cite{Abe:2002rb} are far larger than the leading order NRQCD
theoretical expectations~\cite{Hagiwara:2004pf}.  In this case,
keeping only the leading term in $\alpha_S$ and $v$ in the NRQCD
expansion results in a very rough approximation. In addition, as
opposed to the production of the $J/\psi$ in hadron collisions,
colour-octet channels cannot be invoked to bring theoretical
predictions in agreement with the experimental
measurement~\cite{Liu:2003jj}. Inclusion of the $\alpha_S^3$ (NLO)
corrections to this process has been recently found to reduce the
discrepancy between theory and measurements~\cite{Zhang:2006ay}.

In view of the unexpectedly large measurements for the production of
$J/\psi$ associated with a $c\bar{c}$ quark pair in $e^+e^-$
annihilation, it is natural to wonder whether the corresponding
production pattern could be large in hadroproduction as well.  Besides
offering a new interesting signature, such as $\ell^-\ell^+$ in
association with one or two heavy quark tags, this process contributes
to the $\alpha_S^4$ (NLO) corrections to the inclusive colour-singlet
hadroproduction of $J/ \psi$ and $\Upsilon$. Historically,
these higher-order contributions to the cross section at the Tevatron 
were considered in the fragmentation approximation as a first attempt 
to solve the so-called $\psi'$ anomaly~\cite{Braaten:1994xb,Cacciari:1994dr}.

The purpose of this work is to present and discuss the complete
tree-level calculation for the associated production of $J/ \psi$ and
$\Upsilon$ (commonly noted $\cal Q$) with a heavy-quark pair of the
same flavour. Such associated production has been recently discussed
in the $k_T$ factorisation
formalism~\cite{Baranov:12dh,Baranov:12rz}. Here we consider the
leading order term in the NRQCD expansion in the usual collinear
factorisation scheme.

\begin{figure}[t]
\centerline{\mbox{
\subfigure{\includegraphics[height=2.2cm]{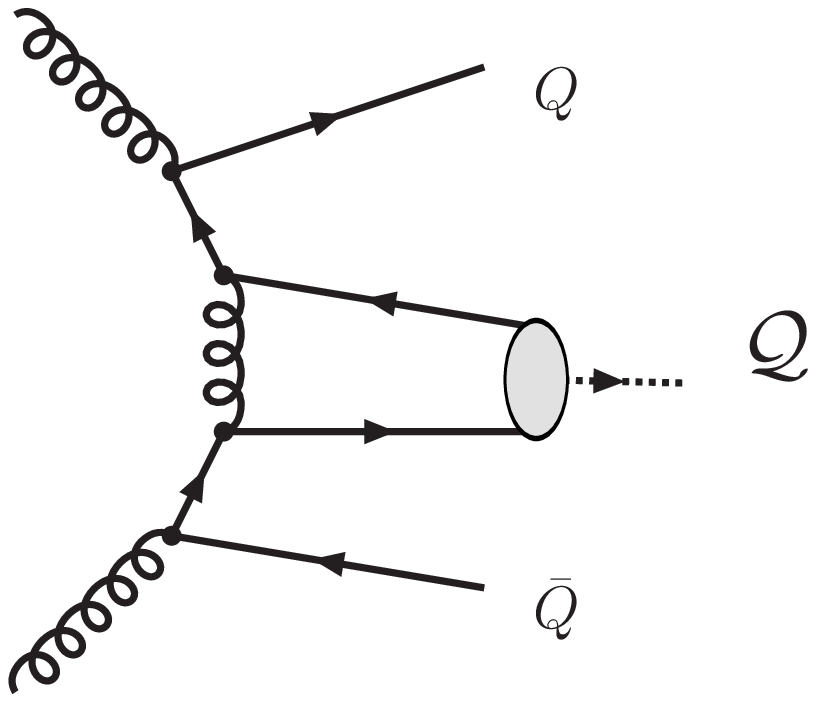}}
\subfigure{\includegraphics[height=2.2cm]{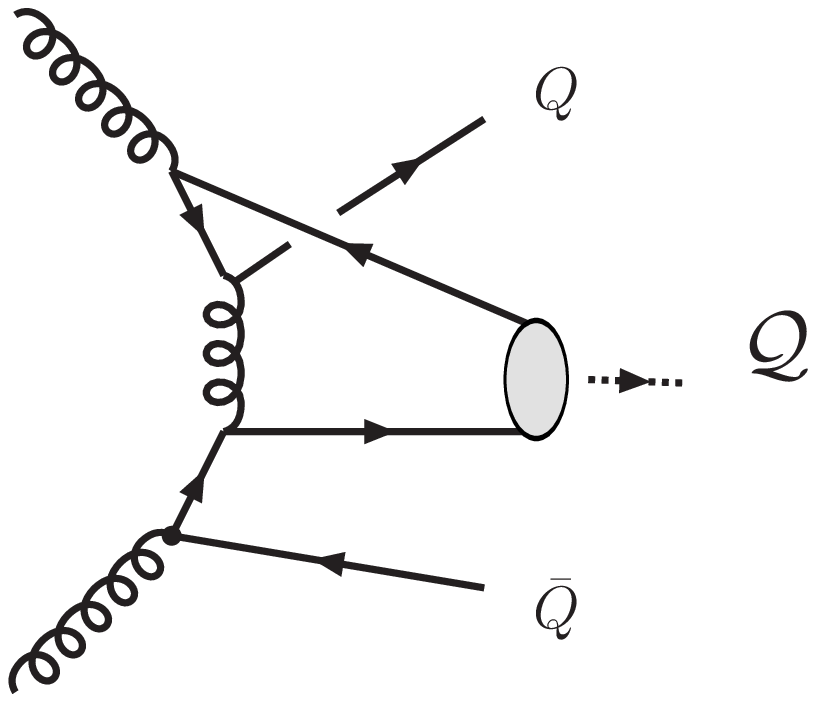}}
\subfigure{\includegraphics[height=2.2cm]{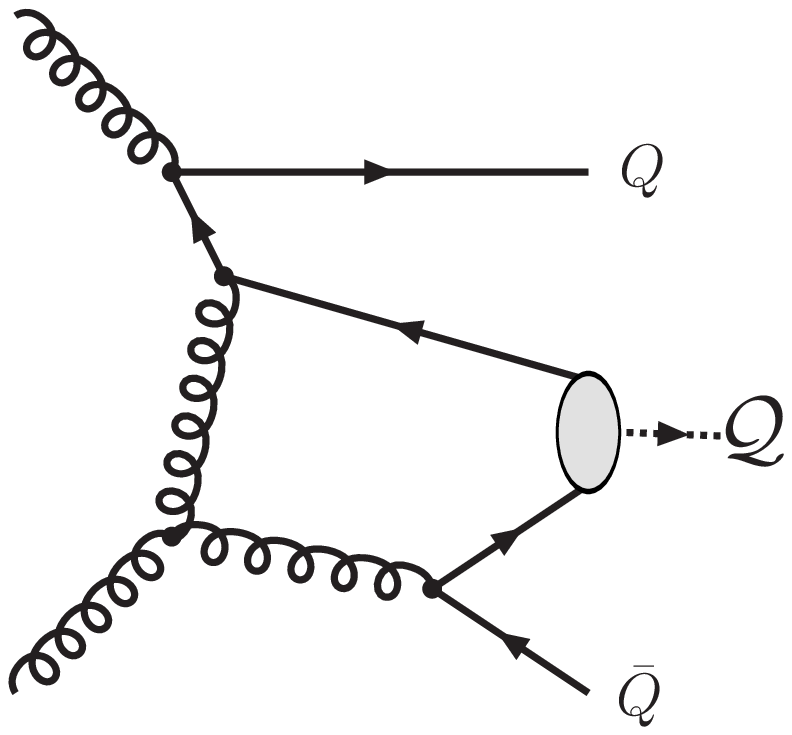}}
\subfigure{\includegraphics[height=2.2cm]{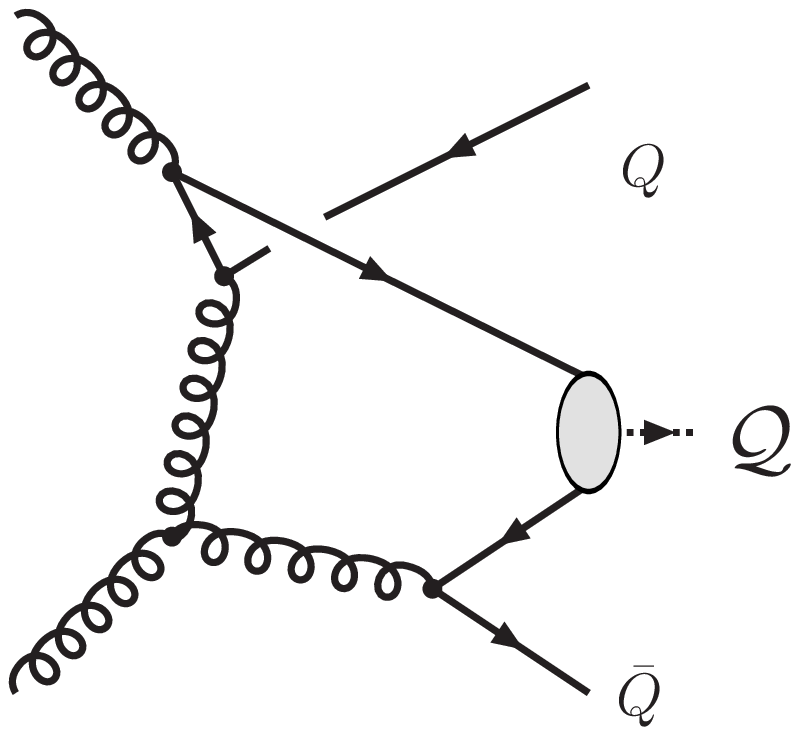}}
\subfigure{\includegraphics[height=2.2cm]{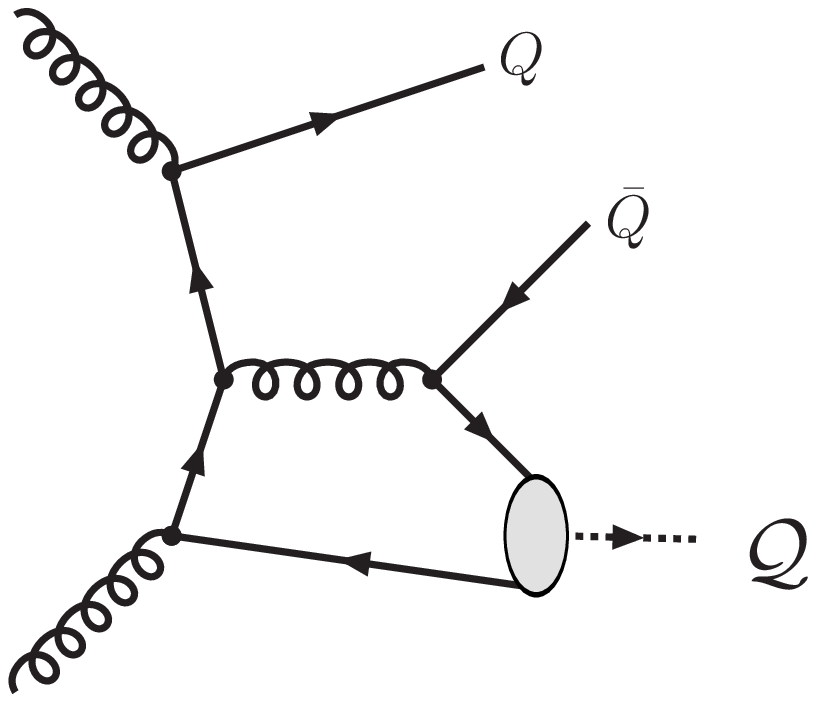}}
}} 
~\\
\centerline{\mbox{
\subfigure{\includegraphics[height=2.2cm]{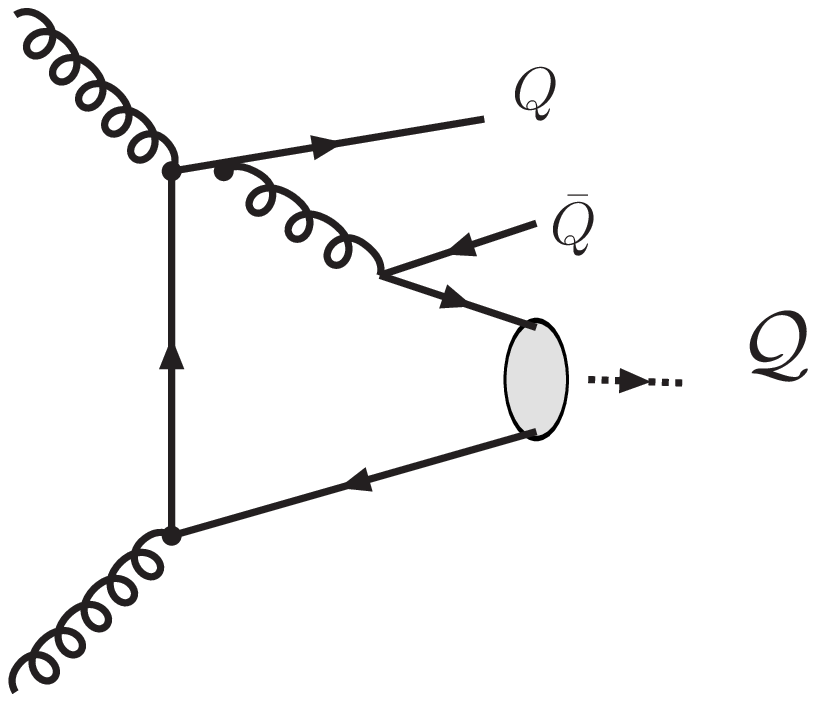}}
\subfigure{\includegraphics[height=2.2cm]{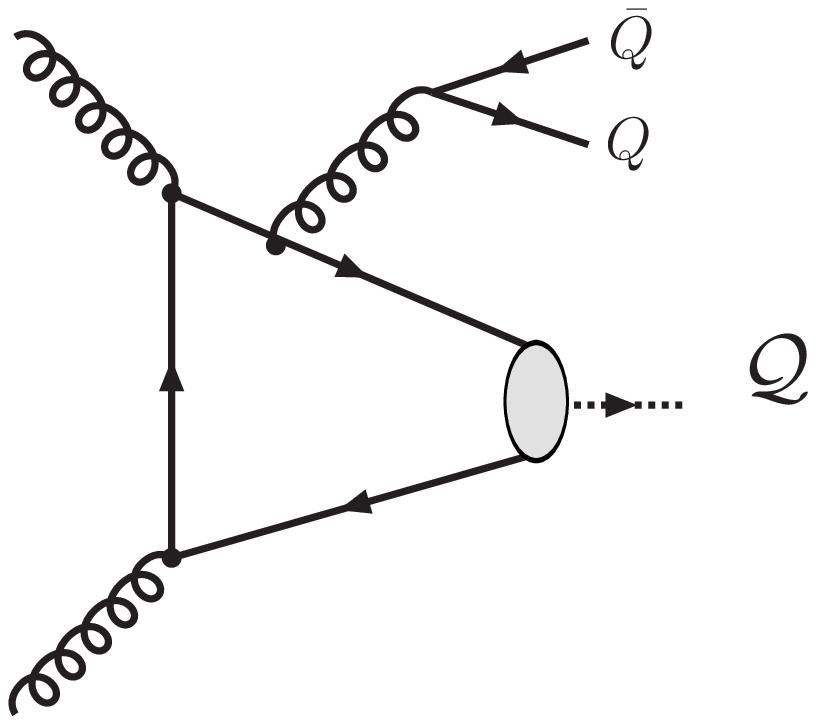}}
\subfigure{\includegraphics[height=2.2cm]{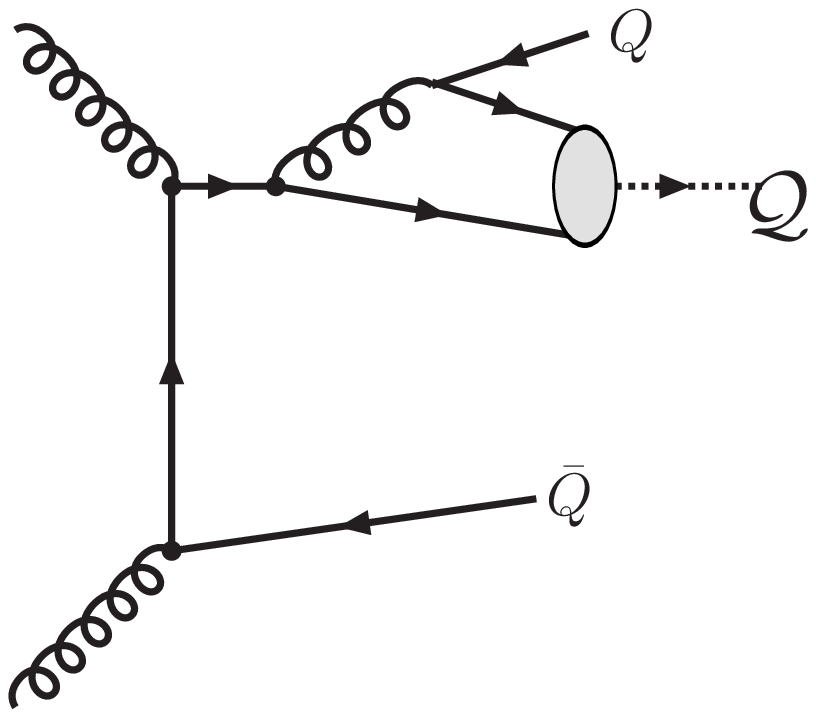}}
\subfigure{\includegraphics[height=2.2cm]{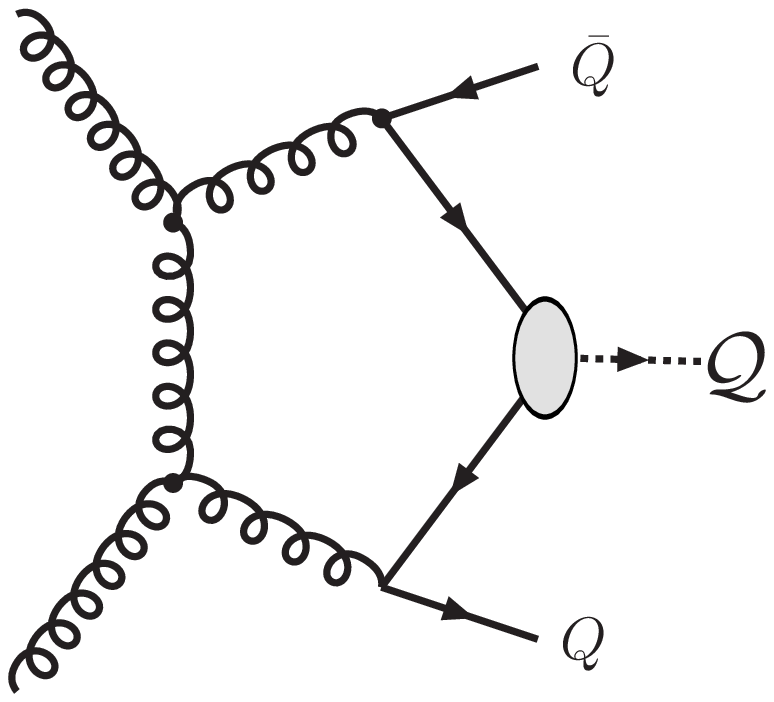}}
}}
\centerline{\mbox{
\subfigure{\includegraphics[height=2.2cm]{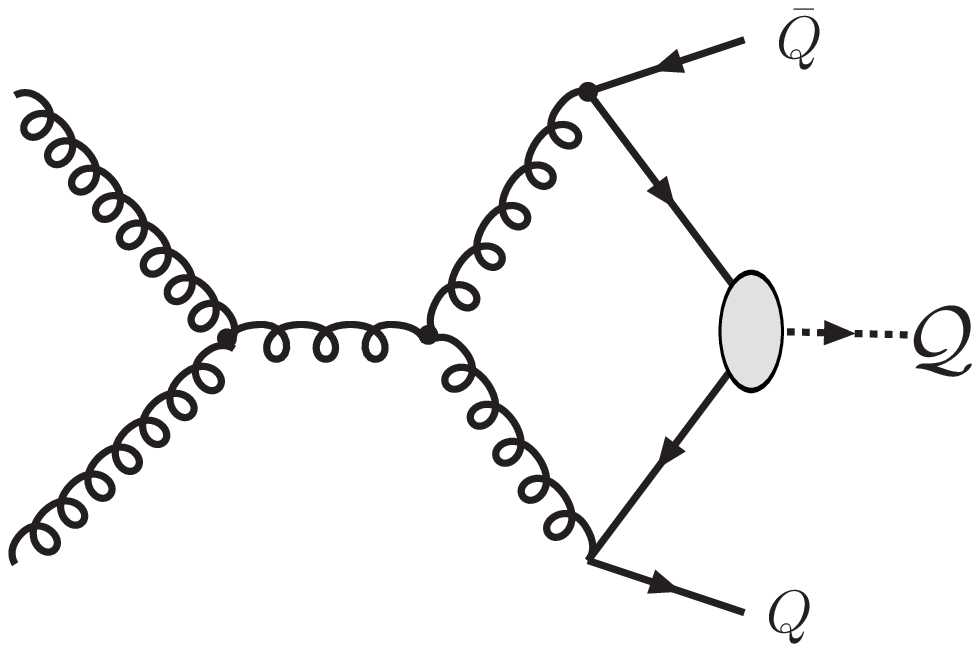}}
\subfigure{\includegraphics[height=2.2cm]{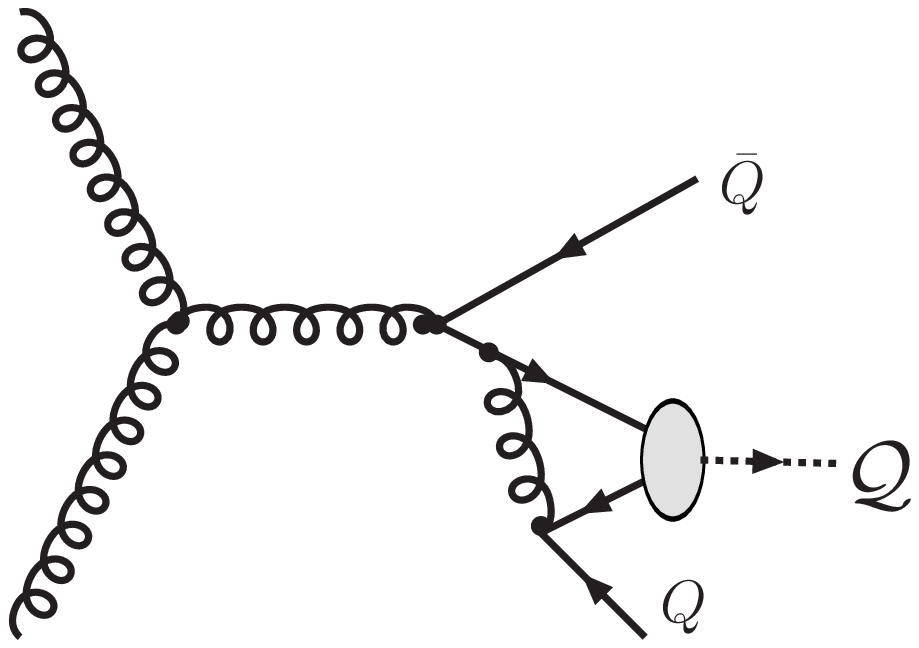}}
\subfigure{\includegraphics[height=2.2cm]{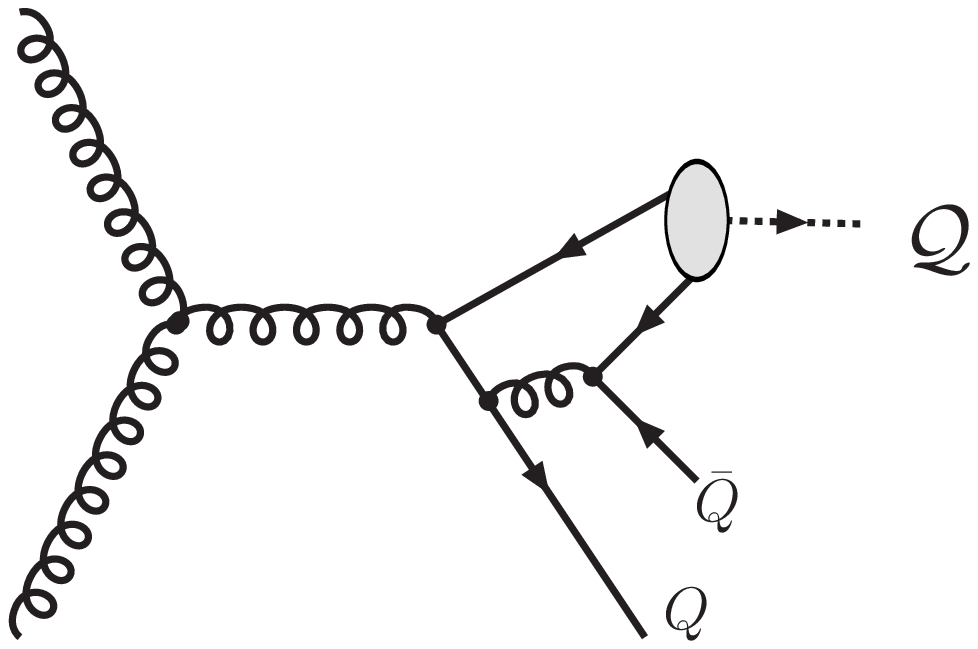}}
\subfigure{\includegraphics[height=2.2cm]{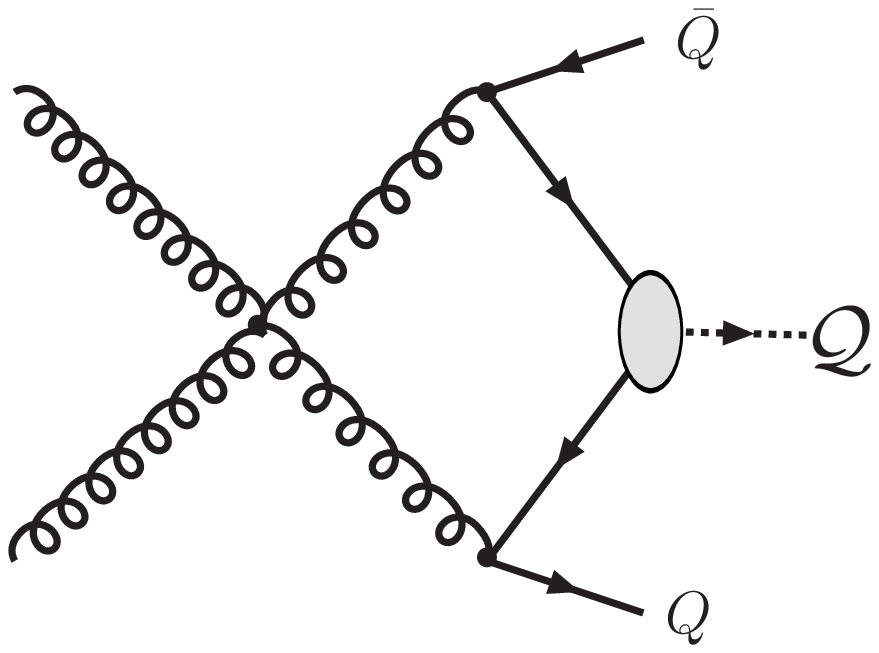}}
}}
\caption{Representative diagrams contributing at LO in $\alpha_S$
to $gg \to {\cal Q} + Q \bar Q $ via a $~^3S_1^{[1]}$ state.}
\label{fig:gg-Qqq}
\end{figure}

This Letter is organised as follows. In Section \ref{sec:LO_ggtoQqq},
we briefly decribe the method used to compute the associated
hadroproduction of a quarkonium with a heavy-quark pair together with 
the results for the Tevatron and the  LHC. In Section
\ref{sec:frag}, we discuss the validity of the fragmentation
approximation $Q \rightarrow \cal Q Q$ at the Tevatron. In Section \ref{sec:polarisation} 
we discuss the polarisation of the quarkonium produced by colour-singlet transitions.
Finally, Section \ref{sec:conclusions} is devoted to the discussion of our results.

\section{The calculation of  $p\bar{p}\to {\cal Q} +  Q\bar Q $}
\label{sec:LO_ggtoQqq}

\subsection{The method}
\label{sec:method}

In this section, we present the numerical method used to compute quarkonium
production amplitudes.  At leading order in
$v$, the invariant amplitude to create
a $^3S_1$ quarkonium ${\cal Q}$ of momentum $P$ and polarisation $\lambda$
 can be expressed as the product of the amplitude to create the corresponding heavy-quark pair, a spin projector $N(\lambda| s_1,s_2)$, and
$R(0)$, the radial wave function at the origin in the configuration
space\footnote{Up to $v^4$ corrections, the relation to the NRQCD
production matrix element is given by $\langle \mathcal{O}_1
\rangle_{J/\psi} =\frac{3N_c}{2 \pi} |R(0)|^2$, where the
normalisation of Ref. \cite{Bodwin:1994jh} is assumed.}, 
namely,~\cite{CSM_hadron,Berger:1980ni}
\eqs{ \label{CSMderiv3} {\cal
M}(gg \to {\cal Q}^\lambda(P)+X)= \sum_{s_1,s_2,i,j}& \frac{N(\lambda|
s_1,s_2)}{ \sqrt{m_Q}} \frac{\delta^{ij}}{\sqrt{N_c}}
\frac{R(0)}{\sqrt{4 \pi}} \times \\ &{\cal M}(gg \to Q_i \bar
Q_j(\mathbf{p}=\mathbf{0};s_1,s_2) + X)\,, }
where $P=p_Q+p_{\bar Q}$,
$p=\frac{p_Q-p_{\bar Q}}{2}$, $s_1$ is the quark spin, $s_2$ the
antiquark one and $\frac{\delta^{ij}}{\sqrt{N_c}}$ is the projector
onto a color-singlet state. In the non-relativistic limit, the spin 
projector factor can be written as
\begin{equation}  \label{pojectionfac}
N(\lambda| s_1,s_2) = 
\frac{\ep^\lambda_{\mu} }{2 \sqrt{2} m_Q } \bar{v} (\frac{\mathbf{P}}{2},s_2) 
\gamma^\mu u (\frac{\mathbf{P}}{2},s_1) \,\, ,
\end{equation}
where $\ep^\lambda_{\mu}$ is the polarisation vector of the quarkonium. 

The amplitude for $gg \to {\cal Q}^\lambda(P) + Q \bar Q$ involves 42
Feynman diagrams and an analytical computation is not practical. We
have therefore opted for using a custom version of
MadGraph~\cite{madgraph}, to generate the amplitudes and perform the
spin projection numerically.  The expression in ~\ce{pojectionfac} can
be evaluated with the help of the HELAS subroutines~\cite{helas}, as done
in Ref.~\cite{Hagiwara:2004pf}. We validated our algorithm to 
generate the amplitudes for quarkonium production and decays by comparing 
with several known analytical results point-by-point in phase space.  
Finally, we use the techniques introduced in Ref.~\cite{madevent} to perform 
the phase-space integration. As an exercise, we have reproduced the cross section and the $P_T$ distribution for $B_c^*$ production at the Tevatron and
found agreement with the results of Ref.~\cite{Berezhnoy:1996ks}.

At hadron colliders a $^3S_1$ quarkonium state is observed through
its decay into leptons. In order to keep 
the spin correlations between the vector-like bound state and the
leptons, we can replace $\ep^\lambda_{\mu}$ in
Eq. (\ref{pojectionfac}) by the leptonic current \eqs{
\frac{\sqrt{3}}{8m_Q \sqrt{\pi}}\bar{u}_{\ell^-}(k_1,\lambda_1)
\gamma_\mu v_{\ell^+}(k_2,\lambda_2) \,. }  
The polarisation of the quarkonium can therefore be determined
by analysing the angular distribution of the leptons. 
Defining $\theta$ as the angle between the $\ell^+$
direction in the quarkonium rest frame and the quarkonium direction in
the laboratory frame, the normalised angular distribution $I(\cos(\theta))$ is
\eqs{ \label{angular_dist} I(\cos \theta) =
\frac{3}{2(\alpha+3)} (1+\alpha \, \cos^2 \theta)\,, } where
the relation between $\alpha$ and the polarisation state of the
quarkonium is \eqs{ \label{def_alpha}
\alpha=\frac{\sigma_T-2\sigma_L}{\sigma_T+2\sigma_L} \,. }

\subsection{Cross sections and results}
\label{sec:results}

As for the case of open charm and bottom cross sections, heavy-quarkonium 
hadroproduction is dominated by gluon fusion.  We have checked that the light-quark initiated
process for $\cal Q + Q\bar{Q}$ production is suppressed by three
orders of magnitude, and thus this contribution is neglected in the following.

In our numerical studies we have used:
\begin{itemize}
\item $|R_{J/\psi}(0)|^2=0.81$ GeV$^3$ and $|R_{\Upsilon(1S)}(0)|^2=6.48$ GeV$^3$;
\item $\mu_0=\sqrt{(4 m_Q)^2+P_T^2}$;
\item Br$(J/\psi \to \mu^+ \mu^- ) = 0.0588$ and Br$(\Upsilon(1S) \to \mu^+ \mu^- ) = 0.0248$
\item $m_c=1.5$ GeV and $m_b=4.75$ GeV;
\item pdf set:  CTEQ6M~\cite{Pumplin:2002vw}.
\end{itemize}

In Fig.2 we show the $P_T$ distributions of $J/\psi$ and
$\Upsilon$ at the Tevatron and the LHC, for both ${\cal Q} +g$ and
${\cal Q}+Q\bar Q$ production.  The theoretical error bands in the
${\cal Q}+Q\bar Q$ curves correspond to the uncertainties from the
renormalisation and factorisation scale (
$\frac{\mu_0}{2} \leq \mu_{f,r} \leq 2 \mu_0$) and the heavy-quark
masses ($m_c=1.5 \pm 0.1$ GeV or $m_b=4.75 \pm 0.05$ GeV), combined in
quadrature.  

We start by considering $J/\psi$ and $\Upsilon$ production at the Tevatron,
where a rapidity cut $|y|<0.6$ is applied.
For the charmonium case, we note that the $P_T$ distribution peaks
at $P_T \simeq m_Q$ and then it starts a quick descrease,
dropping by four orders of magnitude at $P_T \simeq 20$ GeV.  
In fact, at moderate transverse momentum, $P_T \gtrsim  m_Q$,  
${\cal Q}+Q\bar Q$ production has much milder slope
compared to that of ${\cal Q} +g$.
Even though suppressed by $\alpha_S$, ${\cal Q}+Q\bar Q$ 
production starts to be comparable to ${\cal Q} +g$  already for  
$P_T \simeq 4 m_Q$. We verified that the topologies
where the $\cal Q$ is produced by two different quark lines 
always  dominate.
This feature is to be related with the $J/\psi$ production at $e^+e^-$
colliders, where the presence of two extra charmed mesons seems to
indicate that the $J/\psi$ is mostly often created from two different
quark lines as well.  For the $\Upsilon$,
the curve is less steep in the same $P_T$ range due to
the larger value of the $b$-quark mass. 

The results  for the LHC are displayed in \cf{fig:3S1QQ} (c,d). A cut
$|y|<0.5$ has been applied.  Since the rapidity distribution is flat
in this range, this is equivalent to consider the quarkonium
production at zero rapidity. The $P_T$ behaviours for both $J/\psi$ and $\Upsilon$ are
very similar to those at the Tevatron while the normalisation is increased by one order of
magnitude.

\begin{figure}[H]
\centerline{\mbox{
\subfigure[]{\includegraphics[width=6.25cm]{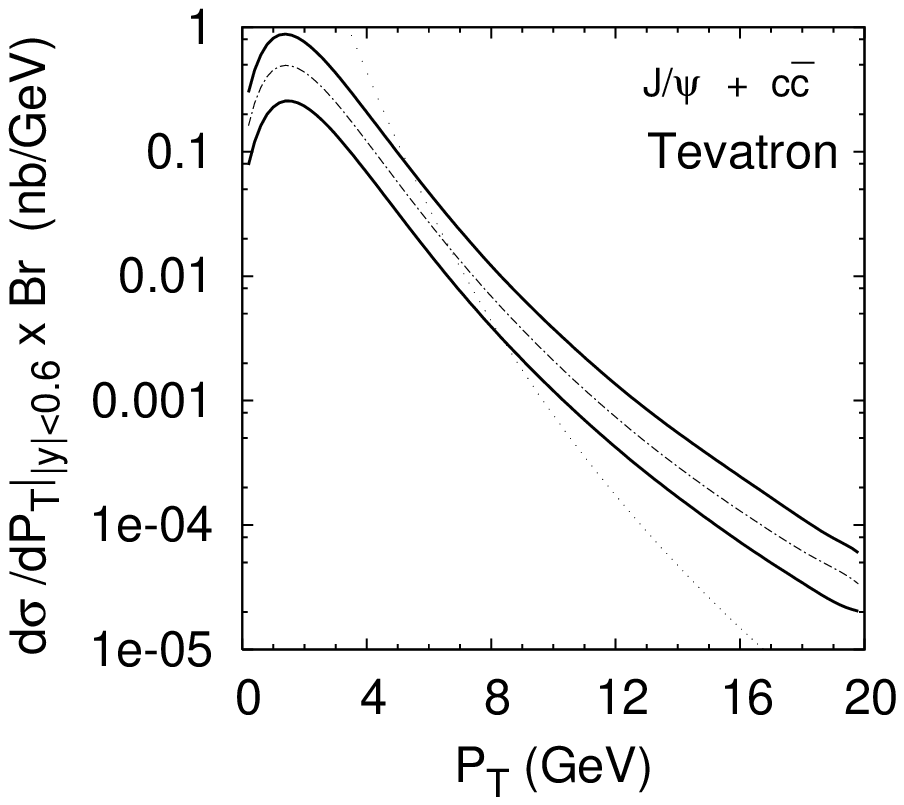}}
\quad
\subfigure[]{\includegraphics[width=6.25cm]{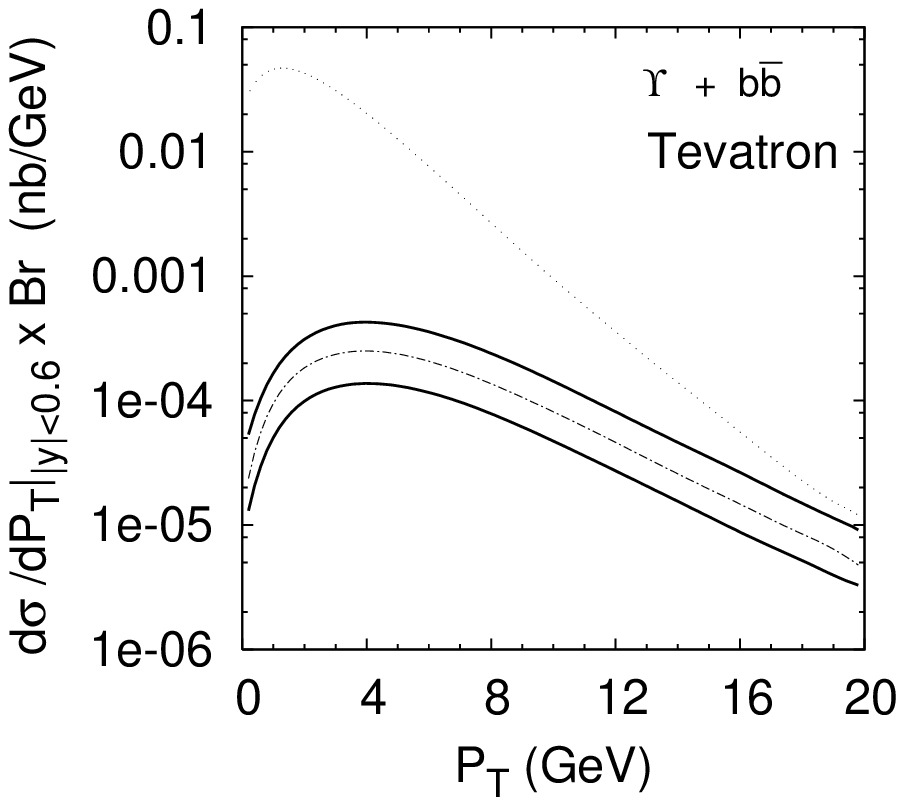}}
}}
\centerline{\mbox{
\subfigure[]{\includegraphics[width=6.25cm]{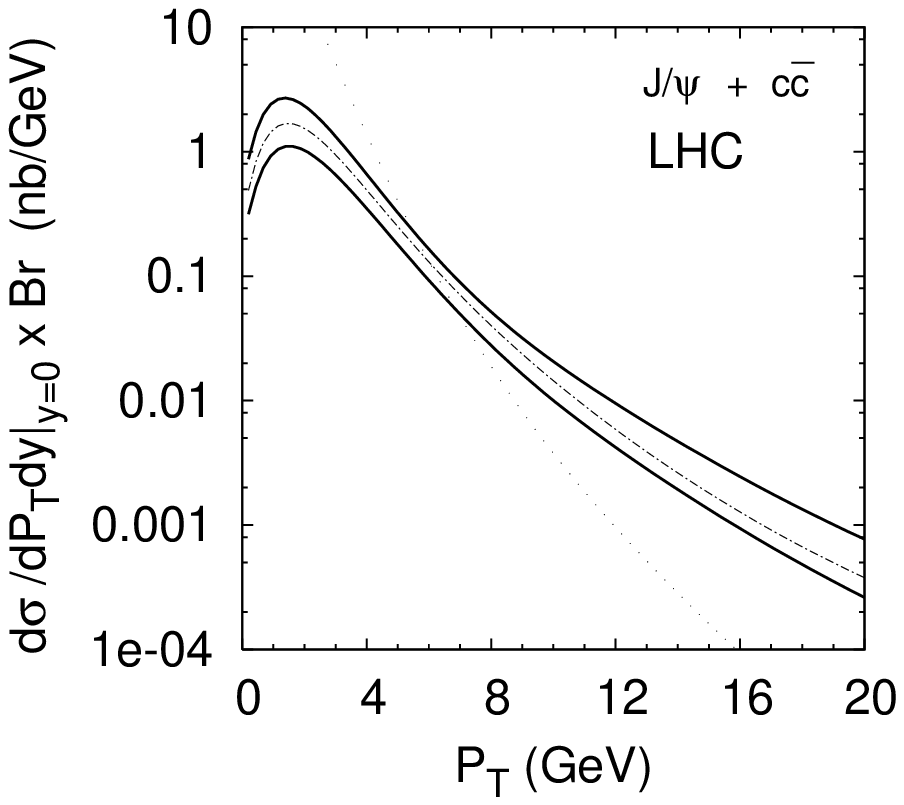}}
\quad
\subfigure[]{\includegraphics[width=6.25cm]{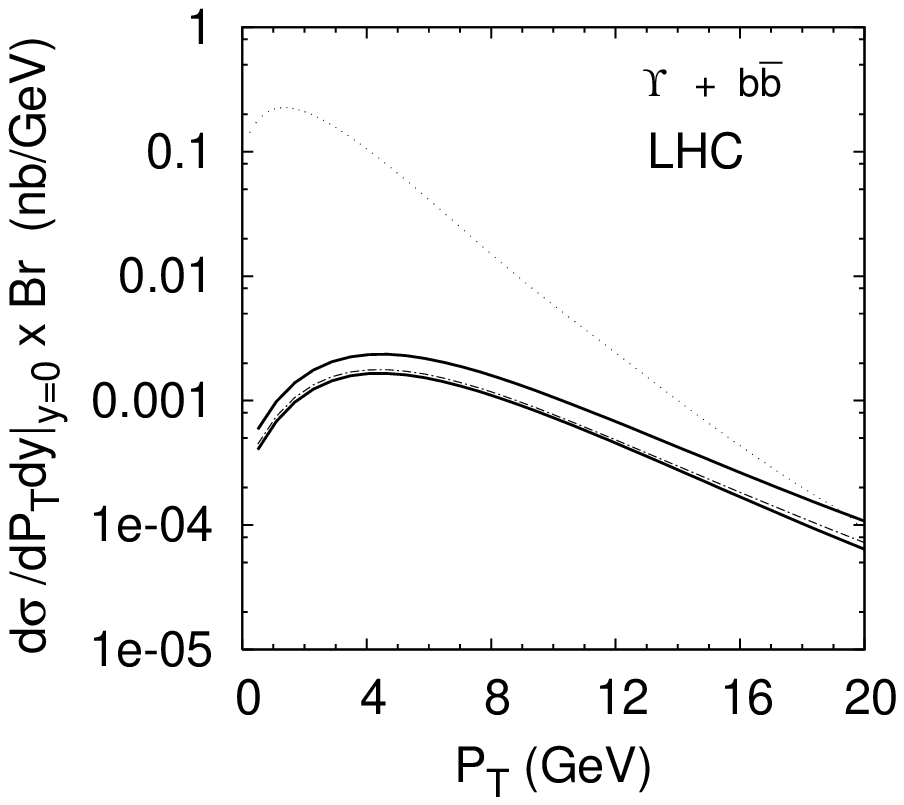}}
}}
\caption{(a) Differential cross section for the process $p\bar{p}\to J/\psi +  c\bar c $  at the 
Tevatron, $\sqrt{s}=1.96$ TeV. For comparison, the 
$P_T$ shape associated to the process $p\bar{p}\to J/\psi +  g $ is also displayed (dotted line).
 (b) Same plot for $\Upsilon + b\bar{b}$ production. (c,d) 
Same plot as (a,b) at the LHC, $\sqrt{s}=14$ TeV.
}
\label{fig:3S1QQ}
\end{figure}

\section{Testing the fragmentation approximation}
\label{sec:frag}

In this section we discuss the range of applicability of the 
fragmentation approximation for the process where a $Q\bar Q$ pair
is produced at high $P_T$ and one of the heavy quarks
fragments into a quarkonium state, $Q \to {\cal Q} Q $.
At sufficiently large $P_T$, it is expected that the
quark-fragmentation contributions would dominate the ${\cal Q} Q\bar Q$ cross section.

\subsection{Heavy-quark fragmentation }

In the fragmentation approximation, the cross section for the
production of a quarkonium $\cal Q$ by gluon fusion via a heavy quark
$Q_i$ fragmentation is given, at all orders in $\alpha_S$, by \eqs{
d \sigma_{{\cal Q}}(P) = \sum_i \int^1_0 \, dz \, d\sigma_{Q_i}(\frac{P}{z},\mu_{frag}) 
D_{Q_i\to{\cal Q}}(z,\mu_{frag}),} where $d\sigma_{Q_i}(\frac{P}{z},\mu_{frag})$ is the differential cross section to
produce an {\it on-shell} heavy quark $Q_i$ with 
momentum $\frac{P}{z}$ and $D_{Q_i\to{\cal Q}}(z,\mu_{frag})$ is the
fragmentation function of $Q_i$ into a quarkonium ${\cal
Q}$. 

The fragmentation scale, $\mu_{frag}$, is usually chosen to avoid
large logarithms of $P_T/\mu_{frag}$ in
$\sigma_{Q_i}(\frac{P}{z},\mu_{frag})$, that is $\mu_{frag}\simeq
P_T$. The summation of the corresponding large logarithms of
$\mu_{frag}/m_Q$ appearing in the fragmentation function can be
 obtained via an evolution equation~\cite{Collins:1981ta,Curci:1980uw,Collins:1981uw}.

The perturbative quark fragmentation function into quarkonium
$\cal{Q}$ via a $^3S_1^{[1]}$ state at the scale $3m_Q$
is~\cite{Braaten:1993mp} \eqs{\label{eq:frag-funct} D_{Q \rightarrow
\cal{Q}}(z, 3 m_Q) =\frac{8 \; \alpha_S^2(2 m_Q)}{27
\pi}\frac{|R(0)|^2}{m_Q^3} {z (1-z)^2 (16 - 32 z + 72 z^2 - 32
z^3 + 5 z^4)\over (2-z)^6}.  }

\subsection{Comparison}

We now compare the results of the full LO cross sections 
for $p\bar{p}\to {\cal Q} + Q \bar Q $ with those calculated in the 
fragmentation approximation. The same set of parameters of 
Section~\ref{sec:results} is employed\footnote{For consistency, the
coupling constant in the fragmentation function is
evaluated at the scale $\sqrt{(4 m_Q)^2 +P_T^2}$. Note that this choice, 
as well as the use of CTEQ6M, leads to smaller cross sections compared those
of Ref.~\cite{Braaten:1993mp}.}. The comparison is shown in
\cf{fig:comp}.  The approximation is lower than the 
full computation in the $P_T$ range accessible at the Tevatron. 
We have verified that for $J/\psi$ production, the two curves still differ by a 
little bit less than $10$\% at $P_T=80$ GeV.  
For the $\Upsilon$, much larger values of $P_T$ are  required (\cf{fig:comp} (b)).

\begin{figure}[h]
\centerline{\mbox{
\subfigure[]{\includegraphics[height=6cm]{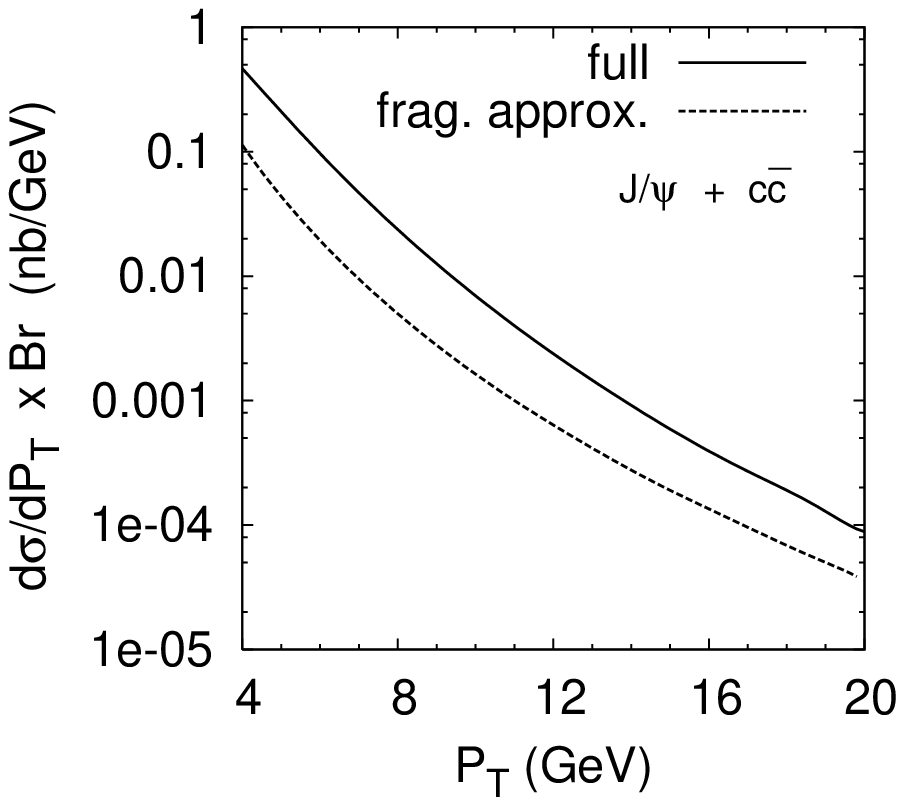}}
\subfigure[]{\includegraphics[height=6cm]{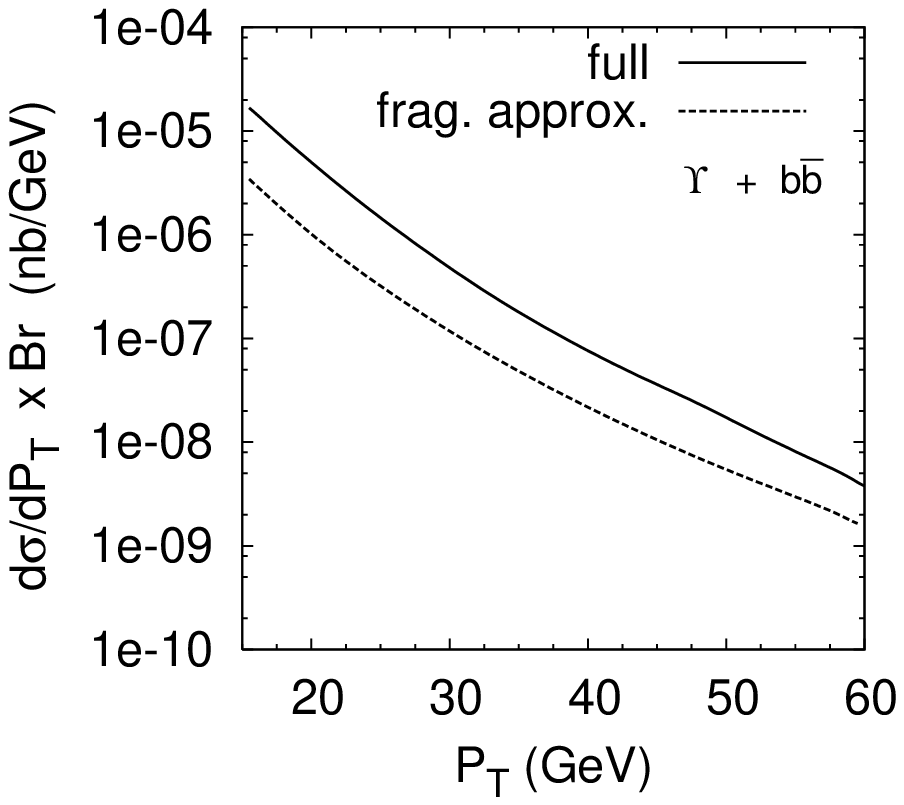}}
}}
\caption{(a) Comparison between the  {\it full} LO cross section for $p\bar p\to J/\psi + c \bar c$ 
and the fragmentation approximation at $\sqrt{s}=1.96$ TeV. 
No cut on rapidity is applied. (b) 
Idem for $p\bar p\to \Upsilon + b \bar b$. }
\label{fig:comp}
\end{figure} 


Therefore, contrary to the common wisdom, in the kinematical region
accessible at the Tevatron, $p\bar{p}\to J/\psi + c \bar c$, which is
an NLO subset of $p\bar{p}\to J/\psi + X$, is not dominated by the
fragmentation contributions.

\section{Polarisation: $gg\to {\cal Q} + Q \bar Q $ vs. $gg\to {\cal Q} + g$}
\label{sec:polarisation}

Let us first review the analytical result for $gg\to {\cal Q} + g$,
which we used as a check of our numerical procedure.  To compute the polarisation
parameter $\alpha$ discussed in Section \ref{sec:method}, it is
sufficient to consider the unpolarised (or total) cross section and
the longitudinal one in the laboratory frame. The parameter $\alpha$
introduced in Section \ref{sec:method} can then be written as
\eqs{\alpha=\frac{\sigma_T-2 \sigma_L}{\sigma_T+2 \sigma_L}=
\frac{\sigma_{tot}-3 \sigma_L}{\sigma_{tot}+ \sigma_L}
.}

The total cross section is calculated from the squared amplitude
 summed over the quarkonium polarisation which
 reads~\cite{CSM_hadron,Gastmans:1987be}
\eqs{\label{eq:M2tot}|{\cal M}|^2= \frac{320 g_S^6 |R(0)|^2 \hat s^2
M_{\cal Q}}{9 \pi} \Big(\frac{(\hat{t}
+ \hat{u})^2+(\hat{s} + \hat{u})^2+(\hat{s} + \hat{t})^2}{(\hat{s} +
  \hat{t})^2 (\hat{s} + \hat{u})^2(\hat{t} + \hat{u})^2}\Big)
,} where $M_{\cal Q}=2 m_Q$ and $\hat{s}$, $\hat{t}$ and $\hat{u}$ are the usual Mandelstam
variables for parton quantities.  To compute the squared amplitude for
longitudinally polarised quarkonia, we need to define the longitudinal
polarisation vector in the laboratory frame:
\eqs{
\ep_3^L(P)&= a_{L} k_1 + b_{L} k_2 +  c_{L} k_4\,,\\
}
with
\eqs{
a_{L}&=\frac{M^2_{\cal Q}(x_1-x_2)-(\hat{t} x_2+\hat{u} x_1)}{M_{\cal Q}N}
\\
b_{L}&=\frac{M^2_{\cal Q}(x_2-x_1)-(\hat{t} x_2+\hat{u} x_1)}{M_{\cal Q}N}
\\
c_{L}&=\frac{-M^2_{\cal Q}(x_1+x_2)+(\hat{t} x_2+\hat{u} x_1)}{M_{\cal Q}N}
\\
N&= \sqrt{((x_1-x_2)^2 M^4_{\cal Q}-2(\hat{u} x_1-\hat{t}x_2)(x_1-x_2)M^2_{\cal Q}+(\hat{t} x_2+\hat{u} x_1)^2)}\,,
}
where $x_1$ and $x_2$ are the usual momentum fractions of the incoming gluons.
From that definition and the Eqs.~(10.505) of Ref.~\cite{Gastmans:1990xh}, we can then easily compute 
\eqs{\label{eq:M2L}|{\cal M}_L|^2=   \frac{640 g_S^6 |R(0)|^2 M^3_{\cal Q} \hat{s} \hat{t} \hat{u} }
{9 \pi (\hat{t}+\hat{u})^2 (\hat{s}+\hat{u})^2 (\hat{s}+\hat{t})^2}\frac{(\hat{t}^2 x_1^2+\hat{s}^2(x_1-x_2)^2+\hat{u}^2 x_2^2)}
{N^2}.}
If $x_1$ and $x_2$ are approximately equal and the masses
can be neglected, it is possible to see from \ce{eq:M2tot} and
\ce{eq:M2L} that the quarkonia will be transversally polarised.
Using the same parameters as for the previous plots, we
find  that $\alpha$ gets rapidly close to one for
$P_T > m_Q$, see \cf{fig:pol-gg3S1g}. Our results agree 
with those of Ref.~\cite{Leibovich:1996pa}.

\begin{figure}[H]
\centerline{\mbox{
\subfigure[]{\includegraphics[width=7cm]{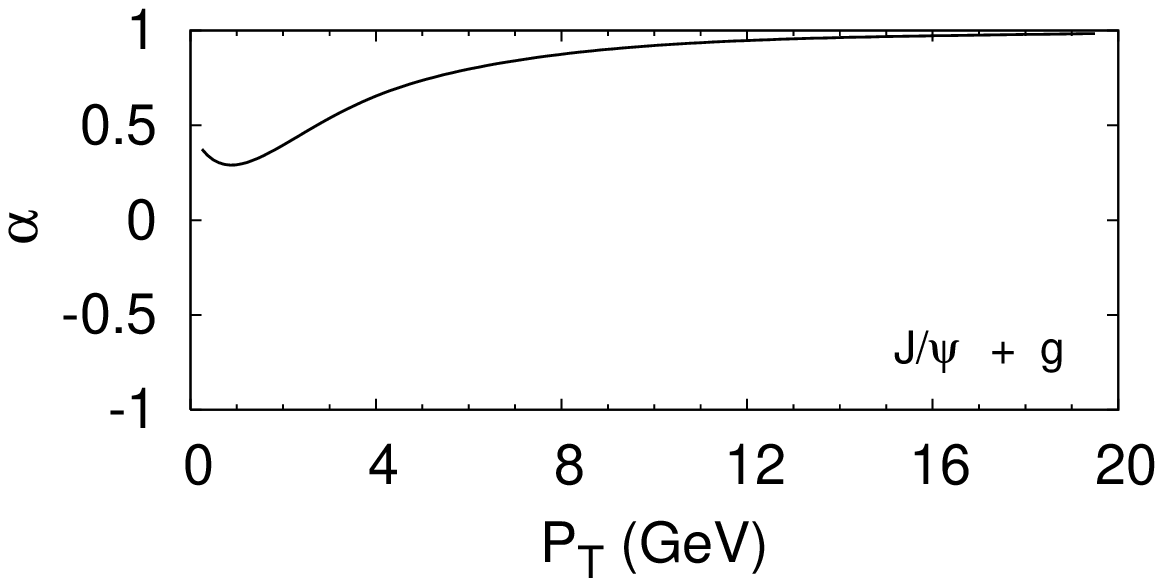}}
\quad
\subfigure[]{\includegraphics[width=7cm]{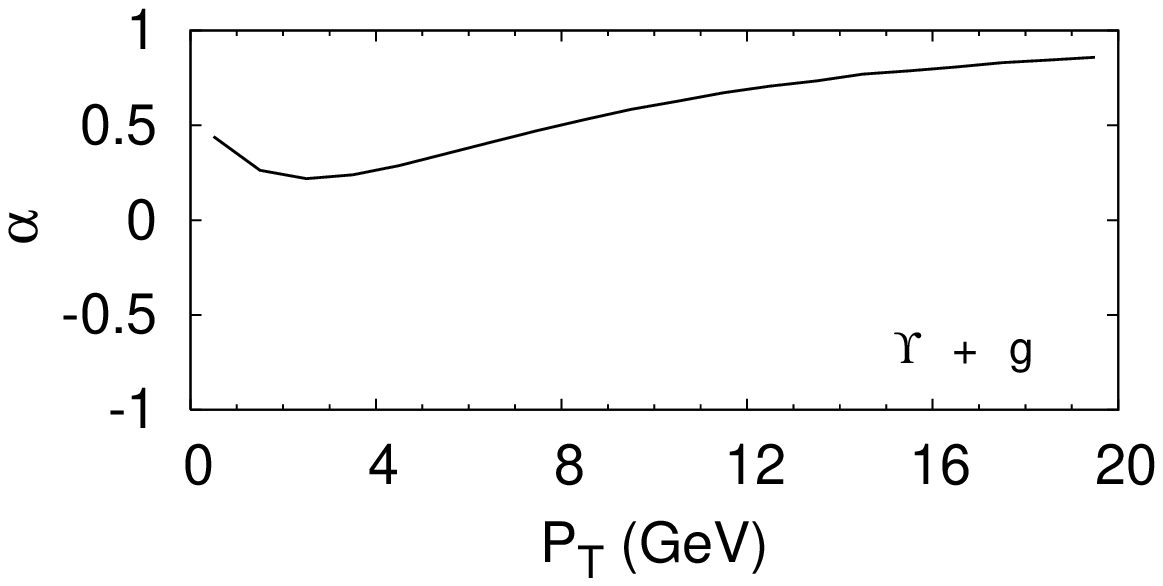}}
\
}}
\caption{Polarisation parameter for $p\bar{p} \rightarrow {\cal Q} + g$ for $J/\psi$ (a)
and $\Upsilon$ (b) at the Tevatron, $\sqrt{s}=1.96$ TeV.}
\label{fig:pol-gg3S1g}
\end{figure} 

Note that this behaviour of $\alpha$ is not expected to be seen
in the data, since the yield from $p\bar p \to {\cal Q} + g$ is not dominant, even when
only colour-singlet channels are considered. For the process 
$gg\to {\cal Q} + Q \bar Q$, $\alpha$ remains close to zero in the whole $P_T$ range.
Since the latter is the one which dominates over the other
colour-singlet contributions, we conclude that the quarkonia not produced by a
colour-octet mechanism will be unpolarised at high $P_T$.

\begin{figure}[H]
\centerline{\mbox{
\subfigure[]{\includegraphics[width=7cm]{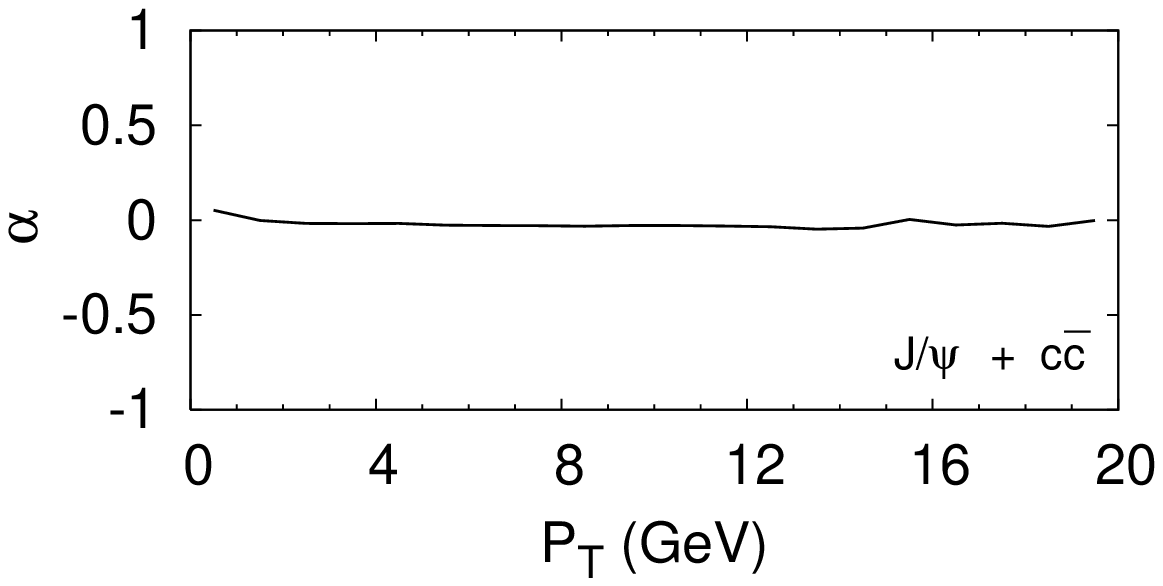}}
\quad
\subfigure[]{\includegraphics[width=7cm]{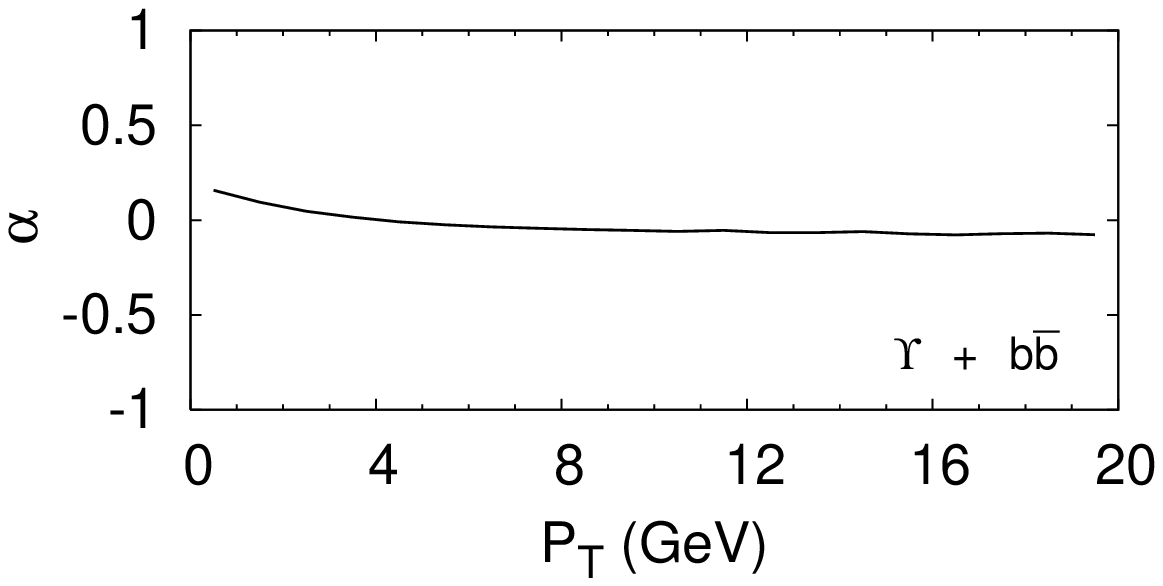}}
\
}}
\caption{Polarisation parameter for $p\bar{p}\to {\cal Q} + Q \bar Q$  for $J/\psi$ (a)
and $\Upsilon$ (b) at the Tevatron, $\sqrt{s}=1.96$ TeV.}
\label{fig:pol-gg3S1QQ}
\end{figure}

\section{Conclusion}
\label{sec:conclusions}

In this Letter, we have presented the tree-level calculation for the
associated production of $J/ \psi$ and $\Upsilon$ with a 
heavy-quark pair. The motivation was twofold.  
First, these processes offer a new interesting signature,
that could be tested experimentally by measuring the fraction of
quarkonium produced with at least one heavy-light quark meson. Second,
they contribute to the $\alpha_S^4$ (NLO) corrections to the inclusive
colour-singlet hadroproduction of $J/\psi$ and $\Upsilon$.

We have found that the quark-fragmentation approximation employed to
describe quarkonium production should not be applied in the range of
transverse momenta reached at the Tevatron and analysed by the CDF
collaboration. The fragmentation approximation actually underestimates
the full contribution of $p \bar{p} \to J/\psi + c \bar{c}$ by more
than a factor four in the region $P_T \simeq 10$ GeV.  This
approximation starts to be justified (within $10 \%$ of error) only
for $P_T \geq 80$ GeV. The same conclusion applies for the $\psi'$.
For $p \bar{p} \to \Upsilon + b \bar{b}$ the fragmentation
approximation is relevant at even higher $P_T$, as expected.  It is
interesting to note that for $J/\psi$ production from $e^+e^-$
annihilation, the fragmentation approximation is only $6 \%$ away from
the full contribution at $\sqrt{s}=50$ GeV \cite{Hagiwara:2004pf}.
One may try to gather an intuitive explanation for such glaring
difference by considering the Feynman diagrams contributing to the two
processes.  In $e^+e^-$ annihilation, there are only four diagrams at
leading order, two of which contribute to the fragmentation
topologies.  On the other hand, the dynamics underlying the
$^3S_1$-quarkonia production at the Tevatron is much more
involved. The number of Feynman diagrams is significantly larger (42
vs. 4) and only a few of them give rise to fragmentation topologies.
As a result, the fragmentation approximation starts to be relevant at
a much higher regime in $P_T$.  A similar situation holds for the
process $\gamma \gamma \to J/\psi c \bar{c}$~\cite{Qiao:2003ba} and
also for the $B_c^*$ hadroproduction, where
the fragmentation approximation is not accurate in the $P_T$ range
explored at the Tevatron~\cite{Chang:1995,Berezhnoy:1996ks}.
We have also shown that the $J/\psi$ or $\Upsilon$ produced in 
association with a heavy-quark pair of the same flavour are {\it unpolarised}.
Further analysis, including combination with the results of
Ref.~\cite{cmt:2007} for $pp \to {\cal Q} + X$ at NLO 
and Ref.~\cite{Petrelli:1997ge}, to compare
with the available data from the Tevatron, is ongoing. 

In conclusion, we look forward to the measurement of the fraction of
events in the $J/\psi$ sample at the Tevatron, with at least one
charmed meson in the final state. Such a measurement could also
provide further insight to the mechanism responsible for inclusive
heavy-quarkonium production at hadron colliders.

\section*{Acknowledgments}

We are thankful to E. Braaten, C.H. Chang, J.R. Cudell,
Yu.L. Kalinovsky, C.F. Qiao, J.W. Qiu, M. Mangano,
A. Meyer and V. Papadimitriou for useful discussions.  
The work of J.P.L is supported by the European I3 program Hadronic Physics, contract
RII3-CT-2004-506078. J.P.L. is also a {\it collaborateur scientifique}
to the University of Li\`ege (Belgium) and thanks the PTF group for
its hospitality.  P.A. is a Research Fellow of the \textit{Fonds
National de la Recherche Scientifique}, Belgium. This work
was supported in part by the Belgian Federal Science Policy (IAP 6/11). 



\end{document}